\documentclass[12pt]{article}
\usepackage{extsizes,graphicx,amssymb,amsmath,amsfonts,mathtext,color,wrapfig}

\def\be{\begin{eqnarray}}
\def\ee{\end{eqnarray}}

\hoffset= - 0.0in         

\begin{document}

\thispagestyle{empty}

\baselineskip14pt

\hfill ITEP/TH-07/15

\vspace{25ex}

\centerline{\Large{Refined Chern-Simons Theory in Genus Two
}}

\vspace{8ex}

\centerline{\large{\emph{S.Arthamonov\footnote{Department of Mathematics, Rutgers, The State University of New Jersey, \\ semeon.artamonov@rutgers.edu; ITEP, Moscow, Russia, artamonov@itep.ru} and Sh.Shakirov\footnote{Department of Mathematics and BCTP, UC Berkeley, USA, \\ shakirov@math.berkeley.edu;
ITEP, Moscow, Russia, shakirov@itep.ru}}}}

\vspace{8ex}

\centerline{ABSTRACT}

\vspace{8ex}

{\footnotesize
Reshetikhin-Turaev (a.k.a. Chern-Simons) TQFT is a functor that associates vector spaces to two-dimensional genus $g$ surfaces and linear operators to automorphisms of surfaces. The purpose of this paper is to demonstrate that there exists a Macdonald $q,t$-deformation -- refinement -- of these operators that preserves the defining relations of the mapping class groups beyond genus 1.\linebreak For this we explicitly construct the refined TQFT representation of the genus 2 mapping class group in the case of rank one TQFT. This is a direct generalization of the original genus 1 construction of arXiv:1105.5117, opening a question if it extends to any genus. Our construction is built upon a $q,t$-deformation of the square of $q$-6j symbol of $U_q(sl_2)$, which we define using the Macdonald version of Fourier duality. This allows to compute the refined Jones polynomial for arbitrary knots in genus 2. In contrast with genus 1, the refined Jones polynomial in genus 2 does not appear to agree with the Poincare polynomial of the triply graded HOMFLY knot homology.
}

\vspace{8ex}

\emph{}

\pagebreak

\section*{Introduction}

Do Chern-Simons TQFT representations of mapping class groups of surfaces have non-trivial deformations? In the case of a torus, it is known \cite{Kirillov, AS} that the answer is positive. Since these representations ultimately determine the TQFT knot invariants, as explained in \cite{AS}, this implies existence of a deformation -- often called refinement -- of the HOMFLY polynomials of torus knots. \cite{AS} observed that refined torus knot invariants agree with the homological knot invariants -- namely, the superpolynomials of \cite{superpoly}, the Poincare polynomials for the triply graded knot homology -- for all torus knots (colored by symmetric or antisymmetric representations). This observation was especially interesting since homological invariants of knots \cite{Khovanov, KhovanovRozansky, KhovanovRozanskyStrings} are generally computationally harder \cite{BarNatan} than TQFT invariants, so the observation of \cite{AS} led to an alternative, more accessible, way to study torus knot homology.

A natural question is how far-going the deformation of TQFT representations, described in \cite{Kirillov, AS}, actually is. There is an ongoing debate in the mathematics and physics community whether it can be extended beyond genus 1, or not. There are arguments both for and against such extension. In this paper, we hope to give convincing evidence that the deformation exists in genus 2, and is related to Macdonald polynomials as directly as in genus 1. This raises a question if this deformation can be similarly carried over in genus 3 and higher, possibly resulting in a full-scale Chern-Simons-Macdonald TQFT. The algebraic approach that we choose in this paper seems to be well-suited to answer this question, and we plan to continue investigating this question in genus 3 and higher.

The genus 2 construction that we suggest shares all features of the genus 1 construction of \cite{AS}, except one. What appears to break down is the striking close relation to homological Poincare polynomials. This was to be expected in the light of the conjecture of \cite{AS} that the refined TQFT computes an index on knot homologies, which accidentally happens to coincide with the Poincare polynomial for simple enough knots and representations. Already in genus 1, if one replaces torus knots by torus links with more than one connected component, or if one replaces the symmetric coloring representations with arbitrary Young diagrams, the literal equality between refined and Poincare polynomials no longer holds true. What happens is that, when one looks at the refined TQFT in increasing generality, signs inevitably start appearing in the coefficients of refined TQFT invariants. This could not happen for the actual Poincare polynomial, but is totally expectable from an index, which is, after all, an Euler characteristic w.r.t. one of the gradings of knot homology.

It appears that generalization to genus 2 knots makes the situation generic enough so that the coincidence with the Poincare polynomial is almost never reached -- unless the knot is actually a torus knot and the above-discussed conditions on the coloring representations are met. It is enough to look at the simplest examples of twist knots ($4_1,6_1,8_1,\ldots$) in s. 6 below to see that they are very different from those of \cite{superpoly,superpoly2}. This, unfortunately, seems to imply that there is little to compare to on the knot homology side. One can only check agreement with Jones polynomials and topological invariance (the latter is a non-trivial check, that we report for a number of interesting knots below).  At the same time, the very existence of refined Chern-Simons in genus 2 suggests existence of extra grading(s) in knot homology, in addition to those already known. If/once these extra gradings are defined, the index conjecture of \cite{AS} for refined Chern-Simons invariants could be checked.

We expect our results to agree with the doubly affine Hecke algebra (DAHA) approach to deformed knot invariants \cite{DAHA0, DAHA1, DAHA2}. While DAHA computations for most genus 2 knots are not available yet, some of them can be computed using a generalization of DAHA described in \cite{CherednikPi}: for example, we were able to confirm the matching for the $6_1$ knot \cite{CherednikLetter}. For more complicated genus 2 knots, however, it does not seem that the generalization of \cite{CherednikPi} is sufficient. Our results seem to suggest that spherical DAHA (a.k.a. the elliptic Hall algebra) admits a genus 2 generalization, generated by knot operators of refined Chern-Simons theory on a genus 2 surface. This will be studied elsewhere.

\paragraph{}From the Reshetikhin-Turaev algebraic viewpoint on TQFT \cite{Reshetikhin, ReshetikhinKirillov, ReshetikhinTuraev, TuraevBlueBook} based on representation theory of the quantum group $U_q(sl_2)$, the present paper relies upon a curious fact: while the q-6j symbols of $U_q(sl_2)$ (and associated fundamental identities such as the pentagon and Yang-Baxter equations) do not seem to admit nice Macdonald deformations, \emph{their squares do}:

\begin{align}
\left\{\left\{\begin{array}{ccc} j_{12} & j_{13} & j_{23} \\ j_{34} & j_{24} & j_{14} \end{array} \right\}\right\}_{q,t} = \left\{\begin{array}{ccc} j_{12} & j_{13} & j_{23} \\ j_{34} & j_{24} & j_{14} \end{array} \right\}_q^2 \ + \ O(q - t)
\end{align}
\smallskip\\
Since the q-6j symbols enter the genus 2 representations only in the squared form, this is enough for the purposes of present paper. The object in the l.h.s. of this equality is an interesting new quantity, which we define and discuss in certain detail in this paper. It is an intriguing question what exactly is the representation theory meaning of this deformation. This observation can also have important consequences for the full refined TQFT, if it exists: it suggests that refined Chern-Simons theory is less local, than usual Chern-Simons theory, since some quantities (the squares of q-6j symbols) that used to be broken up into elementary constituents (the individual q-6j symbols) no longer do so.

The same idea echoes in a different (though related) TQFT, the Turaev-Viro \cite{TuraevViro} a.k.a. BF theory, where the q-6j symbol is an elementary building block of a 3-manifold invariant -- a local weight associated to a single tetrahedron of an arbitrary triangulation. The square of the q-6j symbol is then the weight associated with the simplest triangulation of a 3-sphere into two tetrahedra. The fact that the weight of the whole triangulation admits a deformation, but the local weight of a single tetrahedron does not, might suggest a non-local Macdonald deformation of Turaev-Viro theory. This interesting possibility also needs to be investigated.

\pagebreak

\begin{figure}[t]
\begin{center}
\emph{} \hspace{1ex} \includegraphics[width=0.35\textwidth]{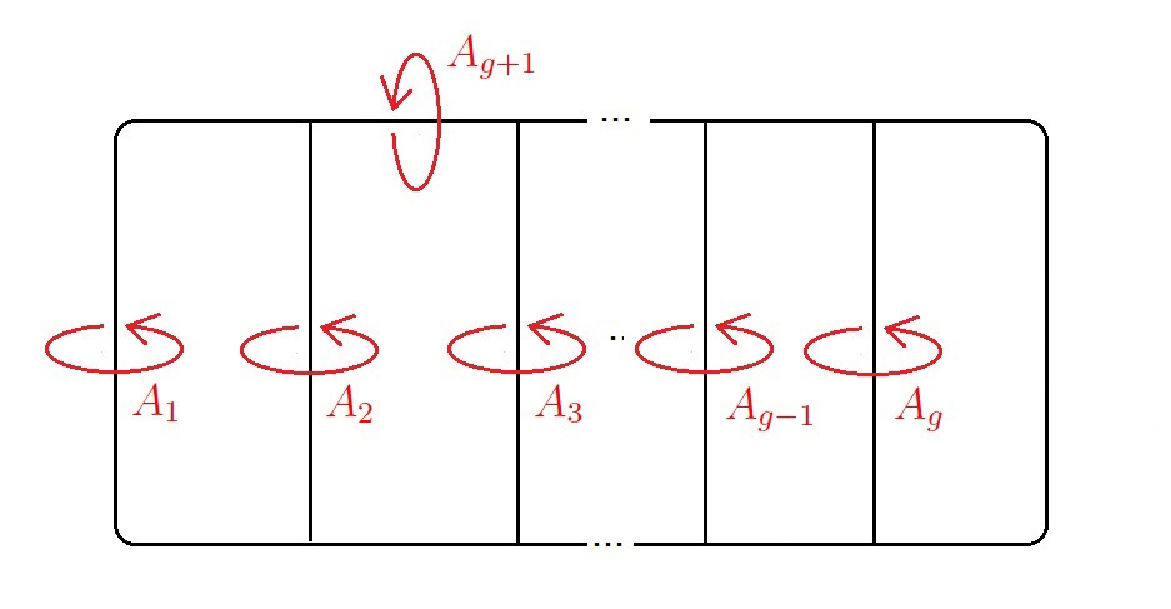}
\end{center}
\caption{The $g+1$ Dehn twists around the A-cycles.}
\label{Atwists}
\end{figure}

\begin{figure}[t]
\begin{center}
\includegraphics[width=0.3\textwidth]{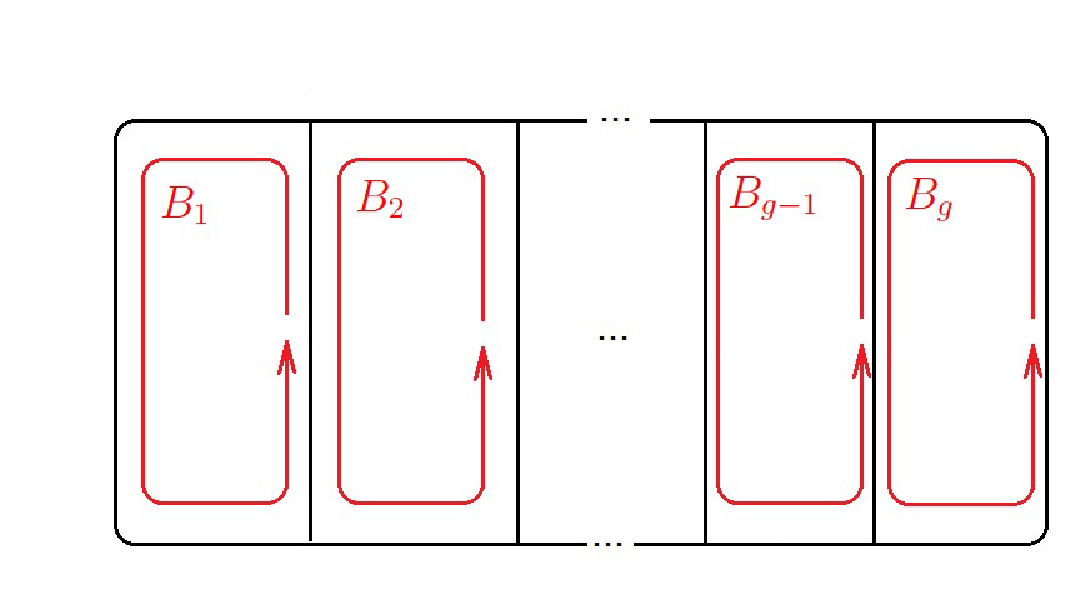}
\end{center}
\caption{The $g$ Dehn twists around the B-cycles.}
\label{Btwists}
\end{figure}

\section{TQFT representations of mapping class groups}

It is well-known \cite{Humphries} that the mapping class group of a genus $g$ closed oriented two-dimensional surface is generated by $2g+1$ Dehn twists along the $A$- and $B$-cycles, shown on Fig.\ref{Atwists} and Fig.\ref{Btwists}, resp., that satisfy algebraic relations \cite{Wajnryb}. These relations can be divided into three types: the degree 2 and 3 relations of a braid group,

\begin{align}
A_n A_m = A_m A_n, \ \ \ \forall \ n,m
\end{align}
\begin{align}
B_n B_m = B_m B_n, \ \ \ \forall \ n,m
\end{align}
\begin{align}
A_n B_m = B_m A_n, \ \ \ \forall \ n,m \ \ \mbox{such that} \ \ {\mathit i}( A_n, B_m ) = 0
\end{align}
\begin{align}
A_n B_m A_n = B_m A_n B_m, \ \ \ \forall \ \ n,m \ \ \mbox{such that} \ \ {\mathit i}( A_n, B_m ) = 1
\end{align}
\smallskip\\
where $i$ is the intersection form, and more exotic higher degree relations, that reflect the difference between mapping class groups and braid groups. One could say that a mapping class group is a braid group with additional higher degree relations. We do not write these additional relations here in full generality, one can easily find them in \cite{Wajnryb}. In the case of genus $g=2$ these additional relations become especially simple and we present a complete set of them in eq. (\ref{RelationsExtra}).

\emph{}

\vspace{3ex}

\begin{figure}[h!]
\begin{center}
\includegraphics[width=0.6\textwidth]{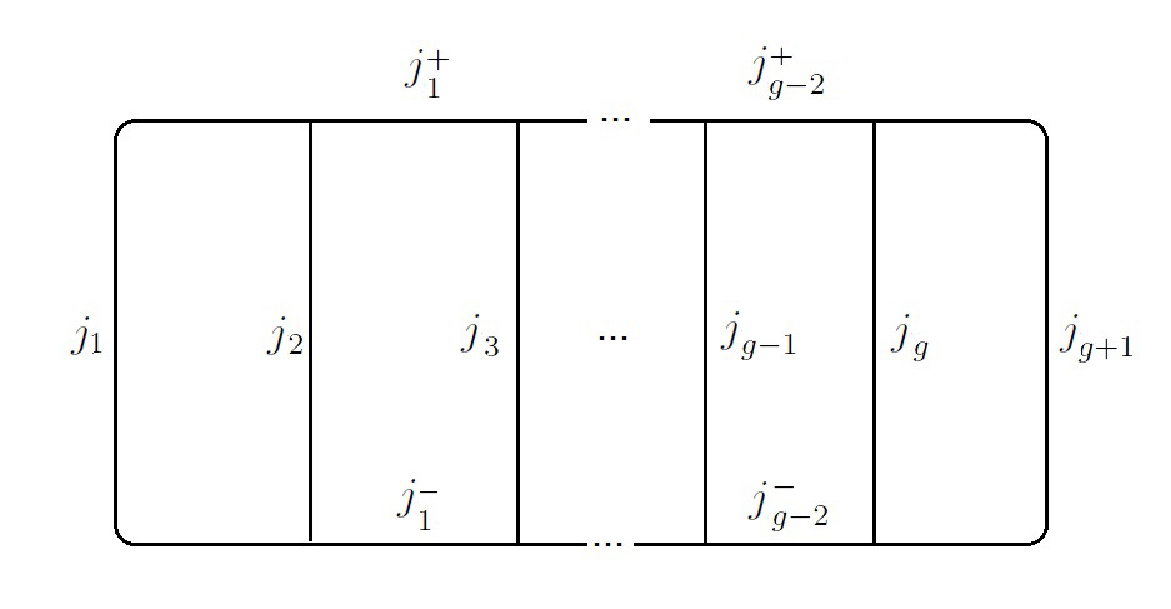}
\end{center}
\caption{Basis vectors in the TQFT vector space, associated to a genus $g$ surface.}
\label{basis-g}
\end{figure}

Rank one, level $K$ Chern-Simons \cite{Witten, ReshetikhinTuraev, TuraevBlueBook} TQFT \cite{Atiyah} is a functor that associates to that surface a vector space, spanned by vectors labeled as on Fig.\ref{basis-g}., where $j_1, \ldots, j_{g+1}$ and $j^\pm_1, \ldots, j^\pm_{g-2}$ are integers in $0, \ldots, K$ such that whenever a triple $(j, j^{\prime}, j^{\prime \prime})$ meets at a vertex, they satisfy the so-called admissibility condition

\begin{align}
| j^{\prime} - j^{\prime \prime}| \leq j \leq j^{\prime} + j^{\prime \prime}, \ \ \ j + j^{\prime} + j^{\prime \prime} = \mbox{ even number } \leq 2K
\end{align}
\smallskip\\
It also associates linear maps to bordisms \cite{Atiyah}; in particular, this implies that the mapping class group of every surface is represented on its vector space by linear operators. To completely describe these representations, it suffices to describe the matrix elements of the generators $A_1, \ldots, A_{g+1}; B_1, \ldots, B_g$. This can be done using any formalism for Chern-Simons TQFT: either by representation theory of the quantum group $U_q(sl_2)$ \cite{Reshetikhin, ReshetikhinKirillov, ReshetikhinTuraev, TuraevBlueBook} or equivalently by skein theory \cite{Masbaum1, Masbaum2, Roberts, ACampo}.

\section{Unrefined TQFT representations for $g = 1,2$}

In this paper, we focus specifically on the cases of $g = 1$ and $g = 2$. This is enough to demonstrate that TQFT representations admit Macdonald deformations beyond the torus case. In these cases the matrix elements of the TQFT representation can be actually expressed in a simple closed form, which is straightforward to prove using either of the methods of \cite{ReshetikhinTuraev, TuraevBlueBook} or \cite{Masbaum1, Masbaum2, Roberts, ACampo}. This form is suggestive of Macdonald deformations. We first describe this closed form, and then give a Macdonald deformation of it.

\pagebreak

\begin{figure}[h!]
\begin{center}
\includegraphics[width=0.15\textwidth]{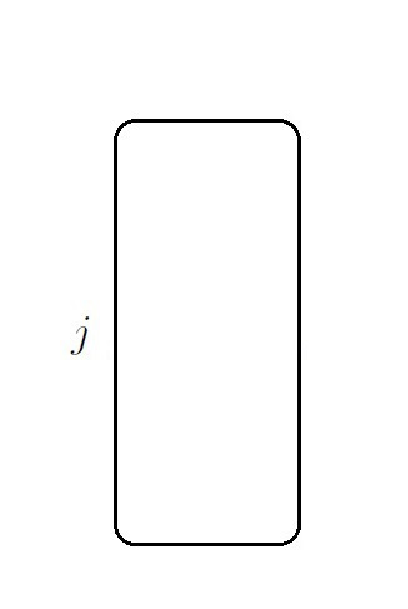} \ \ \ \includegraphics[width=0.2\textwidth]{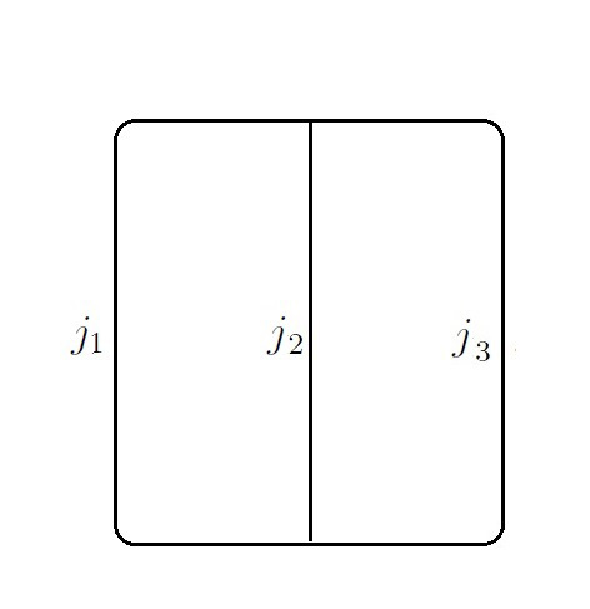}
\end{center}
\caption{Basis vectors in the TQFT vector space, cases $g = 1$ and $g = 2$.}
\label{basis-12}
\end{figure}

\paragraph{For genus 1,}the basis vectors are labeled by a single integer, $0 \leq j \leq K$, as on Fig.\ref{basis-12}. Let us denote that basis vector $| j \rangle$. There are two generators, $A$ and $B$, the representations of which are given \cite{Witten} by the following formulas: for $q = e^{\frac{2 \pi i}{K + 2}}$,

\begin{align}
\langle i | \ A \ | j \rangle = q^{\ j^2/4 + j/2} \ \delta_{i j}
\end{align}
\begin{align}
\langle i | \ A B A \ | j \rangle = \big[(i+1)(j+1)\big], \ \ \ [x] \equiv \dfrac{q^{x/2} - q^{-x/2}}{q^{1/2} - q^{-1/2}}
\label{UnrefinedS}
\end{align}
\smallskip\\
There is a single relation, $ABA = BAB$. The elements $S = ABA$ and $T = A^{-1}$ are often called the modular $S$- and $T$-matrices, because the relations they satisfy closely resemble the $SL(2,{\mathbb Z})$ relations: $S^4 = 1$ and $(ST)^3 = {\rm const} \cdot 1$, with the only difference being an unimportant constant that can be removed by rescaling $T$.

\paragraph{For genus 2,}the basis vectors are labeled by triples of integers, $0 \leq j_1,j_2,j_3 \leq K$, as on Fig.\ref{basis-12}, satisfying an admissibility condition. Let us denote that basis vector $| j_1,j_2,j_3 \rangle$. There are five generators, $A_1,A_2,A_3$ and $B_1,B_2$, with representations

\begin{align}
\langle i_1,i_2,i_3 | \ A_{n} \ | j_1,j_2,j_3 \rangle = q^{ \ j_n^2/4 + j_n/2} \ \delta_{i_1 j_1} \ \delta_{i_2 j_2} \ \delta_{i_3 j_3}, \ \ \ n = 1,2,3
\end{align}
{\fontsize{9pt}{0pt}
\begin{align}
\langle i_1,i_2,i_3 | \ B_1 \ | j_1,j_2,j_3 \rangle = \delta_{i_3 j_3} \ [i_1 + 1][i_2 + 1] \ \sum\limits_{s = 0}^{K} \ q^{-s^2/4-s/2} \ [s+1] \ \left\{\begin{array}{ccc} i_3 & i_2 & i_1 \\ s & j_1 & j_2 \end{array} \right\}_q^2
\end{align}
\begin{align}
\langle i_1,i_2,i_3 | \ B_2 \ | j_1,j_2,j_3 \rangle = \delta_{i_1 j_1} \ [i_2 + 1][i_3 + 1] \ \sum\limits_{s = 0}^{K} \ q^{-s^2/4-s/2} \ [s+1] \ \left\{\begin{array}{ccc} i_3 & i_2 & i_1 \\ j_2 & j_3 & s \end{array} \right\}_q^2
\end{align}}
where the quantity in brackets is the q-6j symbol of the Hopf algebra $U_q(sl_2)$, \cite{ReshetikhinKirillov}:

{\fontsize{9pt}{0pt}
\begin{align}
\left\{\begin{array}{ccc} j_{12} & j_{13} & j_{23} \\ j_{34} & j_{24} & j_{14} \end{array} \right\}_q = \sum\limits_{z} \ \dfrac{ (-1)^z [z+1]! }{ [J_{1} - z]! [J_2 - z]! [J_3 - z]! } \ \prod\limits_{1 \leq a < b < c \leq 4} \dfrac{\Delta\big( j_{ab}, j_{ac}, j_{bc} \big)}{ [z - j_{ab}/2 - j_{ac}/2 - j_{bc}/2 ]! }
\label{KRformula}
\end{align}
\begin{align}
[x]! \equiv [1][2] \ldots [x]
\end{align}
\begin{align}
2J_1 = j_{12} + j_{34} + j_{13} + j_{24}, \ 2J_2 = j_{12} + j_{34} + j_{23} + j_{14}, \ 2J_3 = j_{13} + j_{24} + j_{23} + j_{14}
\end{align}
\begin{align}
\Delta_{ijk} = {\cal N}_{ijk} \ \left( \dfrac{[i/2 + j/2 - k/2]![i/2 - j/2 + k/2]![-i/2 + j/2 + k/2]!}{[i/2 + j/2 + k/2 + 1]!} \right)^{\frac{1}{2}}
\end{align}}
\smallskip\\
The coefficients ${\cal N}_{ijk}$ are often called Verlinde coefficients, and in this case\footnote{We will see later that they deform non-trivially in the refined case. See also \cite{AS}.} are very simple: they take values 1 and 0 depending if the triple $(i,j,k)$ is admissible or not, resp.  The generators satisfy the defining relations of the braid group,

\begin{align}
A_1 B_1 A_1 \propto B_1 A_1 B_1, \ A_2 B_1 A_2 \propto B_1 A_2 B_1
\end{align}
\begin{align}
A_2 B_2 A_2 \propto B_2 A_2 B_2, \ A_3 B_2 A_3 \propto B_2 A_3 B_2
\end{align}
\begin{align}
A_1 A_2 \propto A_2 A_1, \ \ \ A_1 B_2 \propto B_2 A_1, \ \ \ A_1 A_3 \propto A_3 A_1
\end{align}
\begin{align}
B_1 B_2 \propto B_2 B_1, \ \ \ B_1 A_3 \propto A_3 B_1, \ \ \ A_2 A_3 \propto A_3 A_2
\end{align}
\smallskip\\
and a few more exotic relations, which we now write explicitly \cite{Wajnryb}

\begin{align}
\nonumber ( A_1 B_1 A_2 )^4 \propto A_3^2, \ \ \ I^6 \propto 1, \ \ \ H^2 \propto 1 \\
\nonumber \\
H A_n \propto A_n H, \ \ \ H B_n \propto B_n H, \ \ \ \ \ \ \forall n
\label{RelationsExtra}
\end{align}
\smallskip\\
with notations $I = A_1 B_1 A_2 B_2 A_3$ and $H = A_3 B_2 A_2 B_1 A_1 A_1 B_1 A_2 B_2 A_3$. Here, $\propto$ means that the matrices are equal up to a scalar multiple: this implies that TQFT representation is only projective, as it is well known to be the case in general \cite{Masbaum2}.

\section{Refined TQFT representation}

There is an expectation that Chern-Simons TQFT representations of mapping class groups admit a one-parameter deformation, which is characterized, in particular, by deforming the $sl_N$ characters a.k.a. the Schur symmetric polynomials

{\fontsize{9pt}{0pt}
\begin{align*}
& \chi_{1}(x_1, \ldots, x_N) = \sum\limits_{i} x_i, \\
\\
& \chi_{2}(x_1, \ldots, x_N) = \sum\limits_{i} x_i^2 + \sum\limits_{i < j} x_i x_j, \\
\\
& \chi_{3}(x_1, \ldots, x_N) = \sum\limits_{i} x_i^3 + \sum\limits_{i < j} x_i^2 x_j + \sum\limits_{i < j < k} x_i x_j x_k, \ \ \ldots
\end{align*}}
\smallskip\\
into the Macdonald polynomials \cite{Macdonald}:

{\fontsize{9pt}{0pt}
\begin{align*}
& M_{1}(x_1, \ldots, x_N) = \sum\limits_{i} x_i, \\
\\
& M_{2}(x_1, \ldots, x_N) = \sum\limits_{i} x_i^2 + \dfrac{(1-q^2)(1-t)}{(1-q)(1-qt)} \sum\limits_{i < j} x_i x_j, \\
\\
& M_{3}(x_1, \ldots, x_N) = \sum\limits_{i} x_i^3 + \dfrac{(1-q^3)(1-t)}{(1-q)(1-q^2t)} \sum\limits_{i < j} x_i^2 x_j + \dfrac{(1-q^2)(1-q^3)(1-t)^2}{(1-q)^2(1-qt)(1-q^2t)} \sum\limits_{i < j < k} x_i x_j x_k , \ \ \ldots
\end{align*}}
\smallskip\\
Macdonald polynomials depend on two parameters $q$ and $t$, where $t = q^{\beta}$ and $\beta \in {\mathbb C}^{\star}$ is the deformation parameter, so that $\beta = 1$ is the undeformed point. These polynomials are especially simple in the case of rank one, i.e. $N = 2$ eigenvalues:

\begin{align}
\chi_{j}(x_1, x_2) = \dfrac{x_1^{j+1} - x_2^{j+1}}{x_1 - x_2}
\end{align}
\smallskip\\
and, similarly,

\begin{align}
M_{j}(x_1, x_2) = \sum\limits_{l = 0}^{j} \ x_1^{j - l} x_2^l \ \prod\limits_{i = 0}^{l-1} \frac{[j - i]}{[j - i + \beta - 1]} \frac{[i + \beta]}{[i + 1]}
\end{align}
\smallskip\\
\textbf{For genus 1}, a deformation of the TQFT representation has been constructed in \cite{AS}. Let us briefly review it here, concentrating on the rank one, i.e. $N = 2$. The vector space of the refined TQFT remains the same, but the matrix elements of the generators $S$ and $T$ (or equivalently $A$ and $B$) deform,

\begin{align}
\langle i | \ T \ | j \rangle \equiv T_i \delta_{ij} \ = \ q^{-j^2/4} t^{-j/2} \ \delta_{i j}
\end{align}
\begin{align}
\langle i | \ S \ | j \rangle \equiv S_{ij} \ = \ S_{00} \ q^{-ij/2} \ g_{i}^{-1} \ M_{i}\big( t^{\frac{1}{2}}, t^{\frac{-1}{2}} \big) M_{j}\big( t^{\frac{1}{2}} q^i, t^{\frac{-1}{2}} \big)
\end{align}
\smallskip\\
where now $q = e^{\frac{2 \pi i}{K + 2 \beta}}, t = q^{\beta} = e^{\frac{2 \pi \beta i}{K + 2 \beta}}$ (or, equivalently, $t = e^{\frac{2 \pi i}{N}} q^{-K/N}$) and

\begin{align}
g_i = \prod\limits_{m = 0}^{i-1} \dfrac{[i - m] [m + 2 \beta]}{[i - m + \beta - 1][m + \beta + 1]}
\end{align}
\smallskip\\
is the quadratic norm of the Macdonald polynomials under a natural orthogonality condition \cite{AS}. These refined operators satisfy the same relations, as the original ones,

\begin{align}
S^2 = 1, \ \ \ (ST)^3 = \mbox{ central }
\end{align}
\smallskip\\
\textbf{For genus 2}, following the same path, we assume that the vector space is undeformed and the basis vectors are still labeled by triples of integers, $0 \leq j_1,j_2,j_3 \leq K$ satisfying an admissibility condition. We suggest the following formulas for the deformed representations of the five generators, $A_1,A_2,A_3$ and $B_1,B_2$:

\begin{align}
\langle i_1,i_2,i_3 | \ A_{\alpha} \ | j_1,j_2,j_3 \rangle = T_{j_{\alpha}}^{-1} \ \delta_{i_1 j_1} \ \delta_{i_2 j_2} \ \delta_{i_3 j_3}, \ \ \ \alpha = 1,2,3
\label{RefA}
\end{align}
{\fontsize{9pt}{0pt}
\begin{align}
\langle i_1,i_2,i_3 | \ B_1 \ | j_1,j_2,j_3 \rangle = \delta_{i_3 j_3} \ \dfrac{ \dim_{q,t}(i_1) \dim_{q,t}(i_2) }{ {\cal N}_{i_1 i_2 i_3} } \ \sum\limits_{s = 0}^{K} \ T_s \ \dim_{q,t}(s) \ \left\{\left\{\begin{array}{ccc} i_3 & i_2 & i_1 \\ s & j_1 & j_2 \end{array} \right\}\right\}_{q,t}
\label{RefB1}
\end{align}
\begin{align}
\langle i_1,i_2,i_3 | \ B_2 \ | j_1,j_2,j_3 \rangle = \delta_{i_1 j_1} \ \dfrac{ \dim_{q,t}(i_2) \dim_{q,t}(i_3) }{ {\cal N}_{i_1 i_2 i_3} } \ \sum\limits_{s = 0}^{K} \ T_s \ \dim_{q,t}(s) \ \left\{\left\{\begin{array}{ccc} i_3 & i_2 & i_1 \\ j_2 & j_3 & s \end{array} \right\}\right\}_{q,t}
\label{RefB2}
\end{align}}
The logic behind this suggestion is simple: each part of the original formula is replaced by its Macdonald counterpart. E.g. $\dim_{q,t}(i) \ = \ S_{0 i} / S_{0 0} \ = \ M_{i}\big( t^{\frac{1}{2}}, t^{\frac{-1}{2}} \big) $ is a $q,t$-deformation of the quantum dimension $[i+1]$ of the $i$-th representation of $U_q(sl_2)$, ${\cal N}_{ijk}$ is a $q,t$-deformation of the Verlinde coefficients\footnote{Note, that coefficients ${\cal N}_{ijk}$ were trivial in the usual TQFT -- either 1 or 0, depending on whether a triple is admissible or not -- but in the refined setting they are not even integers anymore, but rational functions of $q$ and $t$, and the full formula (\ref{Verlinde}) has to be used to describe them.} discussed in \cite{AS},

{\fontsize{9pt}{0pt}{\begin{align}
{\cal N}_{ijk} = \sum\limits_{l = 0}^{K} \dfrac{ S_{i l} S_{j l} S_{k l} }{ g_{l} S_{0 l} } = g_i^{-1} g_j^{-1} \ \prod\limits_{m = 0}^{\frac{i+j-k}{2} - 1} \dfrac{[i - m][j - m][k + m + 2 \beta][m + \beta]}{[i - m + \beta - 1][j - m + \beta - 1][k + m + \beta + 1][m + 1]}
\label{Verlinde}
\end{align}}}
\smallskip\\
and the quantity in brackets is the deformation of the square (!) of the q-6j symbol,

\begin{align}
\left\{\left\{\begin{array}{ccc} j_{12} & j_{13} & j_{23} \\ j_{34} & j_{24} & j_{14} \end{array} \right\}\right\}_{q,t} = \left\{\begin{array}{ccc} j_{12} & j_{13} & j_{23} \\ j_{34} & j_{24} & j_{14} \end{array} \right\}_q^2 \ + \ O(q - t)
\end{align}
\smallskip\\
that we define and describe in the next section. This is the main new ingredient, not present/seen in genus 1, and the central algebraic quantity of the present paper.

\vspace{2ex}

\paragraph{Conjecture I.} Operators (\ref{RefA}), (\ref{RefB1}), (\ref{RefB2}) satisfy

\begin{align*}
A_1 B_1 A_1 \propto B_1 A_1 B_1, \ A_2 B_1 A_2 \propto B_1 A_2 B_1
\end{align*}
\begin{align*}
A_2 B_2 A_2 \propto B_2 A_2 B_2, \ A_3 B_2 A_3 \propto B_2 A_3 B_2
\end{align*}
\begin{align*}
A_1 A_2 \propto A_2 A_1, \ \ \ A_1 B_2 \propto B_2 A_1, \ \ \ A_1 A_3 \propto A_3 A_1
\end{align*}
\begin{align*}
B_1 B_2 \propto B_2 B_1, \ \ \ B_1 A_3 \propto A_3 B_1, \ \ \ A_2 A_3 \propto A_3 A_2
\end{align*}
\begin{align*}
( A_1 B_1 A_2 )^4 \propto A_3^2, \ \ \ I^6 \propto 1, \ \ \ H^2 \propto 1
\end{align*}
\begin{align*}
H A_n \propto A_n H, \ \ \ H B_n \propto B_n H, \ \ \ \ \ \ \forall n
\end{align*}
\smallskip\\
with notations $I = A_1 B_1 A_2 B_2 A_3$ and $H = A_3 B_2 A_2 B_1 A_1 A_1 B_1 A_2 B_2 A_3$ and, again, $\propto$ is used to stress that the representation is projective. While we cannot yet prove the conjecture in full generality, for any given $K > 0$ it is straightforward to prove by computing the matrices and checking the relations directly. We completed this verification for $1 \leq K \leq 8$; the following is the example of $K = 2$.

\vspace{2ex}

\emph{}

\paragraph{Example: K=2.}The basis of the TQFT vector space consists of 10 vectors

\begin{align*}
|0, 0, 0\rangle, |1, 1, 0\rangle, |2, 2, 0\rangle, |1, 0, 1\rangle, |0, 1, 1\rangle, |2, 1, 1\rangle, |1, 2, 1\rangle, |2, 0, 2\rangle, |1, 1, 2\rangle, |0, 2, 2\rangle
\end{align*}
\smallskip\\
The generators are represented by $10 \times 10$ matrices: $B_1$ is represented by
\begin{center}
\includegraphics[width=0.6\textwidth]{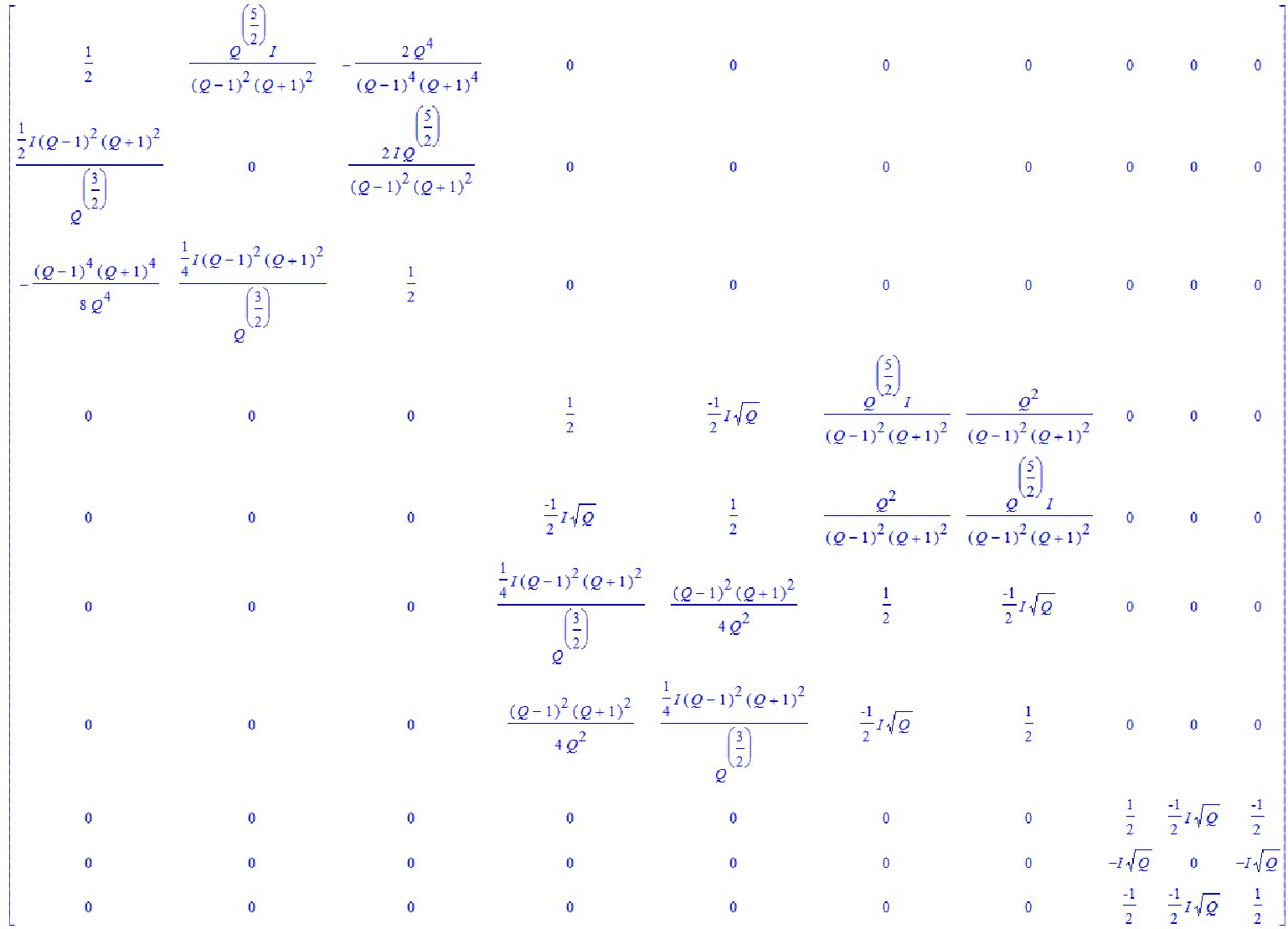}
\end{center}
$B_2$ is represented by
\begin{center}
\includegraphics[width=0.6\textwidth]{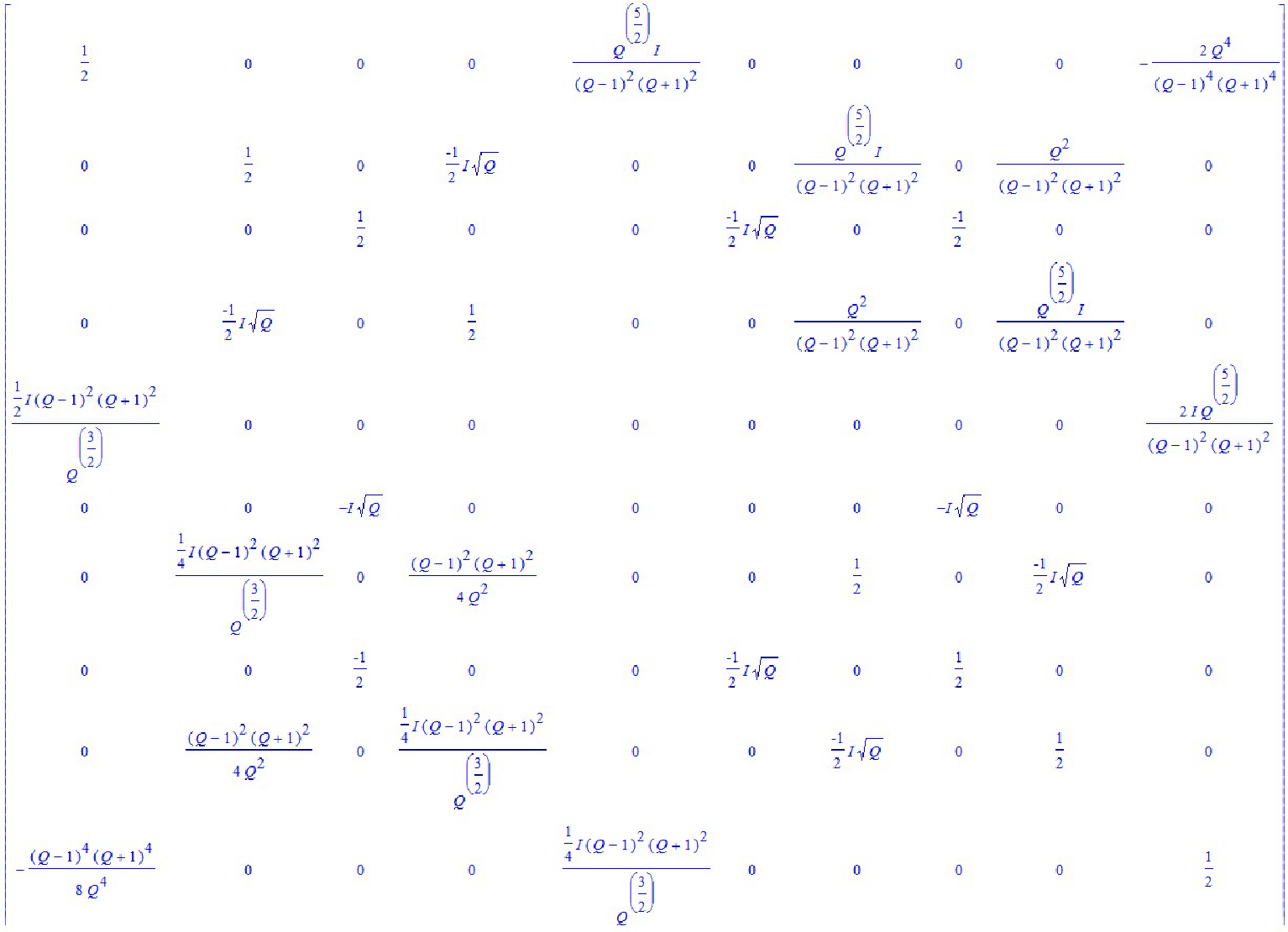}
\end{center}
and $A_1,A_2,A_3$ are trivial. It is straightforward to check that all the relations of the mapping class group are satisfied. Here $Q = e^{\frac{\pi i}{2 \beta + 2}} = \sqrt{q}$ is a refinement parameter, that reduces to the standard TQFT value at $\beta = 1$, that is, $Q = e^{\frac{\pi i}{4}}$, $q = t = e^{\frac{\pi i}{2}}$. It is equally straightforward to produce such matrices for any $K$.\linebreak

\section{The deformation of the q-6j symbol squared}

It appears that the square of the $q$-6j symbol admits a $q,t$-deformation. The definition of this object is the following: it is the unique solution\footnote{We have verified that the solution exists and is unique for $1 \leq K \leq 8$.} to the linear system of equations that we suggest to call the \emph{Macdonald duality} equation:

{\fontsize{9pt}{0pt}
\begin{align}
\left\{\left\{\begin{array}{ccc} j_{12} & j_{13} & j_{23} \\ j_{34} & j_{24} & j_{14} \end{array} \right\}\right\}_{q,t} \ = \ \sum\limits_{ i_{12},i_{13},i_{23},i_{14},i_{24},i_{34} = 0 }^{K} \ \prod\limits_{a < b} S_{j_{ab}, i_{ab}} \ \left\{\left\{\begin{array}{ccc} i_{34} & i_{24} & i_{14} \\ i_{12} & i_{13} & i_{23} \end{array} \right\}\right\}_{q,t}
\label{duality}
\end{align}}
\smallskip\\
that has the same symmetries (24 permutations) and zeroes (if any of the 4 triples are non-admissible) as the standard q-6j symbol. The representation-theory meaning of this quantity, covariant under Macdonald duality, remains to be seen. The equation is the Macdonald analog of the well-known \emph{Fourier duality} of the square of the q-6j symbol, originally found in the Regge quantum gravity literature \cite{Duality1,Duality2}:

{\fontsize{9pt}{0pt}
\begin{align}
\left\{\begin{array}{ccc} j_{12} & j_{13} & j_{23} \\ j_{34} & j_{24} & j_{14} \end{array} \right\}_q^2 \ = \ \sum\limits_{ i_{12},i_{13},i_{23},i_{14},i_{24},i_{34} = 0 }^{K} \ \prod\limits_{a < b} \ S^{(q=t)}_{j_{ab}, i_{ab}} \ \left\{\begin{array}{ccc} i_{34} & i_{24} & i_{14} \\ i_{12} & i_{13} & i_{23} \end{array} \right\}_{q}^2
\end{align}}
\smallskip\\
The reason for this name is that the explicit form (\ref{UnrefinedS}) of the unrefined $S$-matrix looks like a (discrete or difference) Fourier transform. The fact that the refined S-matrix is an analog and a generalization of the Fourier transform, in particular that it is self-dual ($S^2 = 1$), has been discussed in detail in \cite{CherednikFourier}. By solving Macdonald duality, we can compute any desired number of examples. A first few are as follows:

\begin{align}
\left\{\left\{\begin{array}{ccc} 0 & 0 & 0 \\ 0 & 0 & 0 \end{array} \right\}\right\}_{q,t} \equiv 1
\end{align}
\begin{align}
\left\{\left\{\begin{array}{ccc} 0 & 0 & 0 \\ 1 & 1 & 1 \end{array} \right\}\right\}_{q,t} = \dfrac{t^{1/2}(1 - qt)}{(1 + t)^2 (1 - q)}
\end{align}
\begin{align}
\left\{\left\{\begin{array}{ccc} 0 & 1 & 1 \\ 2 & 1 & 1 \end{array} \right\}\right\}_{q,t} = \dfrac{t(1 - qt)^2}{(1-q)^2(1+t)^4}
\end{align}
\begin{align}
\left\{\left\{\begin{array}{ccc} 0 & 2 & 2 \\ 2 & 2 & 2 \end{array} \right\}\right\}_{q,t} = \dfrac{t^{2}(1 - q^2t)^2(1-qt)^4(1-q^2t^2)(1-t)}{(1-q)^3(1+t)^4(1-q^3t)(1-qt^2)^4}
\end{align}
\begin{align}
\left\{\left\{\begin{array}{ccc} 1 & 1 & 2 \\ 2 & 2 & 1 \end{array} \right\}\right\}_{q,t} = \dfrac{t^{3/2}(1-t)(1-qt)^3(1-q^2t^2)}{(1+t)^5(1-q)^3(1-qt^2)^2}
\end{align}
\begin{align}
\nonumber & \left\{\left\{\begin{array}{ccc} 2 & 2 & 2 \\ 2 & 2 & 2 \end{array} \right\}\right\}_{q,t} = \dfrac{t^2(1+qt)(1-t)(1-q^2t)^3(1-qt)^6}{(1+t)^5(1-q)^4(1-q^3t)^3(1-qt^2)^5} \times \emph{} \\
\nonumber \\
& \emph{} \big( \ 1-2 t+q t-q t^2+3 q^2 t-2 q^2 t^2-2 q^3 t+3 q^3 t^2-q^4 t+q^4 t^2-2 q^5 t^2+q^5 t^3 \ \big)
\end{align}
\smallskip\\
and so on. One can see that the non-vanishing quantities with one 0-index are

\begin{align}
\left\{\left\{\begin{array}{ccc} 0 & n & n \\ v & u & u \end{array} \right\}\right\}_{q,t} = \dfrac{ {\cal N}_{n,u,v} }{ \dim_{q,t}(n) \dim_{q,t}(u) }
\label{0index}
\end{align}
\smallskip\\
and the non-vanishing quantities with one 1-index are

{\fontsize{6pt}{0pt}\begin{align}
\left\{ \left\{ \begin{array}{ccc} 1 & n & n+1 \\ v & u+1 & u \end{array} \right\} \right\}_{q,t} = \dfrac{{\cal N}_{n+1,u-1,v}}{ \dim_{q,t}(n+1) \dim_{q,t}(u+1) } \ \dfrac{\left[ \frac{n+u+v}{2} + 1 + \beta \right] \ \left[ \frac{n+u-v}{2} + 1 \right] \ \big[ u + 2 \beta \big] \ \big[ n + 2 \beta \big] }{ \big[ u + \beta \big] \ \big[ u + \beta + 1 \big] \ \big[ n + \beta \big] \ \big[ n + \beta + 1 \big] }
\end{align}}
\smallskip\\
{\fontsize{7pt}{0pt}\begin{align}
\left\{ \left\{ \begin{array}{ccc} 1 & n & n+1 \\ v & u-1 & u \end{array} \right\} \right\}_{q,t} = - \dfrac{{\cal N}_{n+1,u+1,v}}{ \dim_{q,t}(n+1) \dim_{q,t}(u-1) } \ \dfrac{\left[ \frac{u+v-n}{2} + 1 + \beta \right] \ \left[ \frac{u-v-n}{2} + 1 \right] \ \big[ n + 2 \beta \big] }{ \big[ u + 2 \beta - 1 \big] \ \big[ n + \beta \big] \ \big[ n + \beta + 1 \big]}
\end{align}}
\smallskip\\
generalizing the well-known specializations of q-6j symbols. The quantities with indices $2,3,4,\ldots$ can be described with equally explicit formulas, see s. 7.

\paragraph{}Eq. (\ref{0index}) explains, among other things, consistency with the genus 1 case: indeed, genus 2 TQFT has a subsector $i_1 = j_1 = 0$ which looks precisely like a torus, and the B-twist in that subsector reproduces the torus B-twist as expected:

\begin{align}
\sum\limits_{s = 0}^{K} \ T_s \ \dim_{q,t}(s) \ \left\{\left\{\begin{array}{ccc} i & i & 0 \\ j & j & s \end{array} \right\}\right\}_{q,t} \ \propto \ \sum\limits_{s = 0}^{K} \ T_s \ \dim_{q,t}(s) \ {\cal N}_{i j s} = (T S T)_{i j}
\end{align}
\smallskip\\
Here we used the well-known formula for $TST$ in terms of ${\cal N}$ \cite{AS}.

\pagebreak

\begin{figure}[t]
\begin{center}
\includegraphics[width=1.0\textwidth]{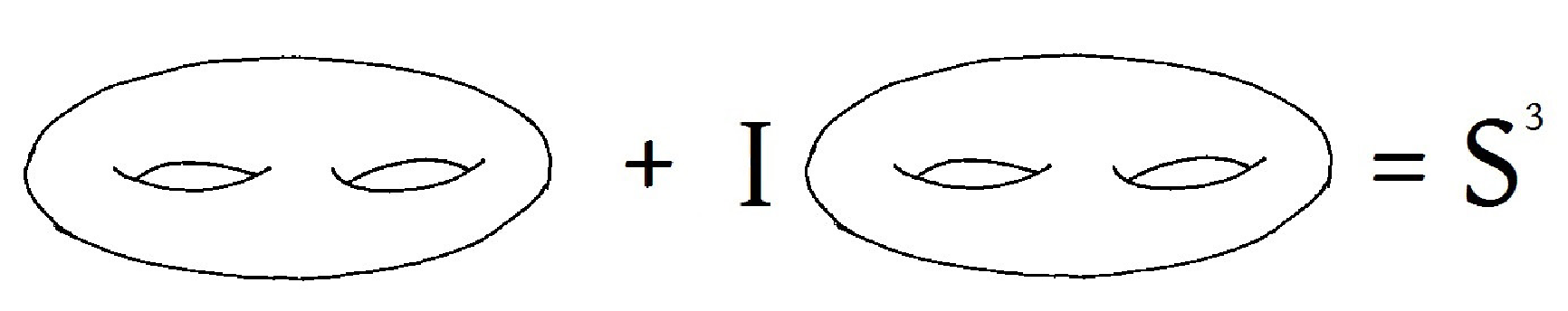}
\end{center}
\caption{Heegaard splitting: gluing an $S^3$ from two genus 2 handlebodies.}
\label{Heegaard}
\end{figure}

\section{Knot invariants}

As explained in \cite{AS}, to compute the TQFT knot invariants, in addition to the representation of the mapping class group one also needs the \emph{knot operators} ${\cal O}_j(K)$, that the TQFT functor associates to the bordisms inserting $K$ colored by representation $j$. We define these operators below. The $j$-colored knot amplitude of $K$ is

\begin{align}
{\cal Z}_j( K ) = \langle 0, 0, 0 | \ I {\cal O}_j(K) \ | 0, 0, 0 \rangle
\end{align}
\smallskip\\
This represents the geometric operation of gluing an $S^3$ from two genus $2$ handlebodies. One first takes a vector $| 0, 0, 0 \rangle$ -- the state corresponding to an empty handlebody -- then acts on it by the knot operator to insert a knot into it, and finally takes a scalar product with another vector $I | 0, 0, 0 \rangle$ to glue in the second handlebody. Note that the boudaries of the handlebodies are not glued identically, as this would not result in an $S^3$; instead, they are glued with the help of an inversion transformation $I = A_1 B_1 A_2 B_2 A_3$ in analogy with the torus case. This way to obtain $S^3$ is called Heegaard splitting \cite{Heegaard}, see Fig.\ref{Heegaard}.

\paragraph{}Based on the computations below, and on the relation to mapping class groups, we propose the following topological invariance conjecture:

\paragraph{Conjecture II.} $|{\cal Z}_j( K )|^2 \equiv {\cal Z}_j( K; q,t ){\cal Z}_j( K; q^{-1},t^{-1} )$ is an invariant of knots.

\paragraph{}Note that, by construction, at $q = t$ the refined Chern-Simons amplitude ${\cal Z}_j( K )$ coincides with the Jones polynomial of $K$. For $q \neq t$ the amplitude ${\cal Z}_j( K )$ provides a refinement of the Jones polynomial. Unlike for the unrefined Jones polynomial, only the norm of the knot amplitude is topologically invariant; as for the phase, a very simple counterexample is presented below. If Conjecture II is true, this new knot invariant does not distinguish mirrors, by construction. However, at this cost it may be better in distinguishing more complicated aspects of knot theory, in particular, the mutant pairs \cite{mutants}. 

\pagebreak

\begin{figure}[h!]
\begin{center}
$
{\color{blue}{{\cal O}^{(1)}_j}} = \begin{array}{c} \includegraphics[width=0.2\textwidth]{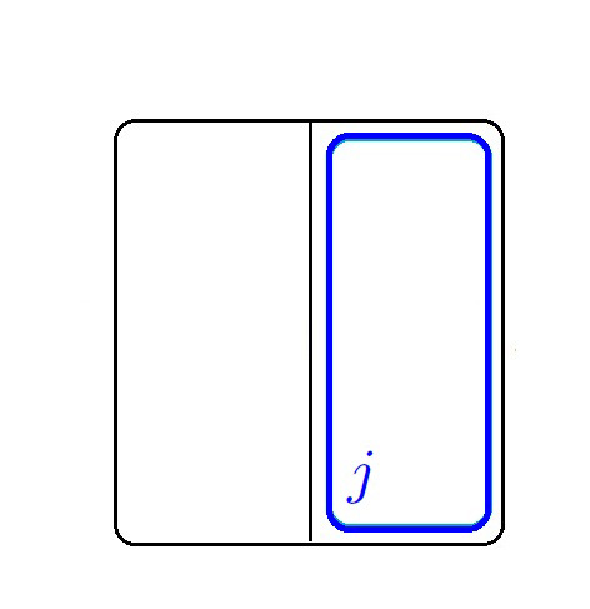} \end{array} \ \ \
{\color{blue}{{\cal O}^{(2)}_j}} = \begin{array}{c} \includegraphics[width=0.2\textwidth]{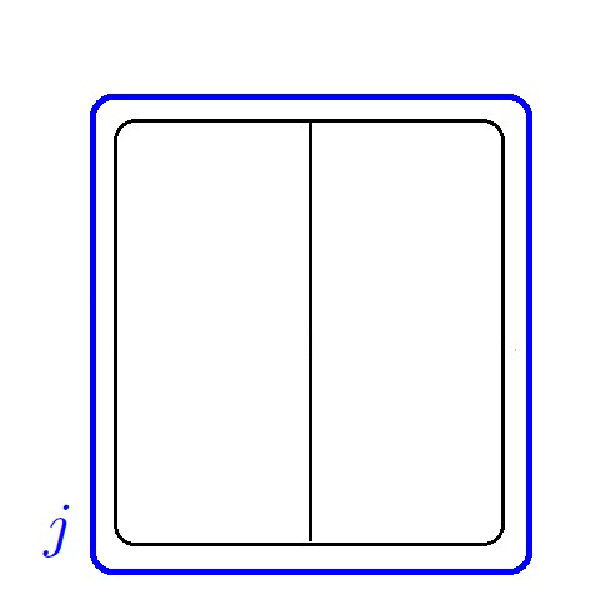} \end{array} \ \ \
{\color{blue}{{\cal O}^{(3)}_j}} = \begin{array}{c} \includegraphics[width=0.2\textwidth]{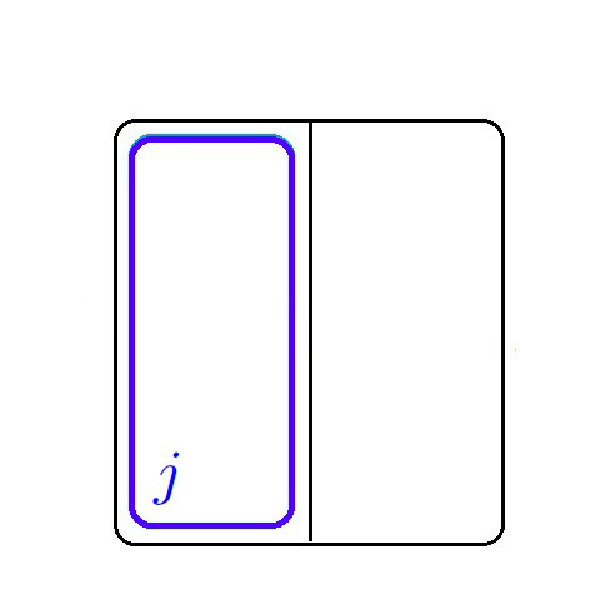} \end{array}
$
\end{center}
\vspace{-2ex}
\caption{Simplest knot operators, corresponding to insertions of unknots.\newline The upper index represents which vertical line out of 3 is left untouched.}
\end{figure}

Let us start with simple knot operators, representing unknots that wind around the first handle (the 3-unknot) the second (the 1-unknot) or both (the 2-unknot). Their matrix elements, computed, say, using the methods of \cite{Masbaum1}, have a form

{\fontsize{10pt}{0pt}
\begin{align}
\langle i_1,i_2,i_3 | \ {\cal O}^{(1)}_j \ | j_1,j_2,j_3 \rangle = \delta_{i_1 j_1} \ [i_2 + 1] [i_3 + 1] \ \left\{\begin{array}{ccc} i_3 & i_2 & i_1 \\ j_2 & j_3 & j \end{array} \right\}_q^2
\end{align}
\begin{align}
\langle i_1,i_2,i_3 | \ {\cal O}^{(2)}_j \ | j_1,j_2,j_3 \rangle = \delta_{i_2 j_2} \ [i_1 + 1] [i_3 + 1] \ \left\{\begin{array}{ccc} i_3 & i_2 & i_1 \\ j_1 & j & j_3 \end{array} \right\}_q^2
\end{align}
\begin{align}
\langle i_1,i_2,i_3 | \ {\cal O}^{(3)}_j \ | j_1,j_2,j_3 \rangle = \delta_{i_3 j_3} \ [i_1 + 1] [i_2 + 1] \ \left\{\begin{array}{ccc} i_3 & i_2 & i_1 \\ j & j_1 & j_2 \end{array} \right\}_q^2
\end{align}}
\smallskip\\
We propose the following Macdonald deformation of these formulas:

{\fontsize{10pt}{0pt}
\begin{align}
\langle i_1,i_2,i_3 | \ {\cal O}^{(1)}_j \ | j_1,j_2,j_3 \rangle = \delta_{i_1 j_1} \ g_j \ \dfrac{ \dim_{q,t}(i_2) \dim_{q,t}(i_3) }{ {\cal N}_{i_1 i_2 i_3} } \ \left\{\left\{\begin{array}{ccc} i_3 & i_2 & i_1 \\ j_2 & j_3 & j \end{array} \right\}\right\}_{q,t}
\end{align}
\begin{align}
\langle i_1,i_2,i_3 | \ {\cal O}^{(2)}_j \ | j_1,j_2,j_3 \rangle = \delta_{i_2 j_2} \ g_j \ \dfrac{ \dim_{q,t}(i_1) \dim_{q,t}(i_3) }{ {\cal N}_{i_1 i_2 i_3} } \ \left\{\left\{\begin{array}{ccc} i_3 & i_2 & i_1 \\ j_1 & j & j_3 \end{array} \right\}\right\}_{q,t}
\end{align}
\begin{align}
\langle i_1,i_2,i_3 | \ {\cal O}^{(3)}_j \ | j_1,j_2,j_3 \rangle = \delta_{i_3 j_3} \ g_j \ \dfrac{ \dim_{q,t}(i_1) \dim_{q,t}(i_2) }{ {\cal N}_{i_1 i_2 i_3} } \ \left\{\left\{\begin{array}{ccc} i_3 & i_2 & i_1 \\ j & j_1 & j_2 \end{array} \right\}\right\}_{q,t}
\end{align}}
\smallskip\\
Once an unknot is inserted, one can use the action of the mapping class group to wind it into something non-trivial. Fig.\ref{figureeight} illustrates how this is done, starting from a 2-unknot, then doing transformations $B_1^{-1} A_1$ and $B_2^{-1} A_3$ to wind it around the handles, and finally $A_2$ to complete the knot. What one obtains is a figure eight knot, a.k.a. $4_1$. This gives an explicit formula for the knot operator, that inserts the $4_1$ knot, colored by representation $j$:

\begin{align}
{\cal O}_j( 4_1 ) = U \ {\cal O}^{(2)}_j \ U^{-1},  \ \ \ U = A_2 B_2^{-1} A_3 B_1^{-1} A_1 
\end{align}
\smallskip\\
More generally, quite a large family of genus 2 pretzel knots can be obtained by further acting on the figure eight knot by the three $A$-twist operators:


\begin{figure}[t]
\begin{center}
\includegraphics[width=1.0\textwidth]{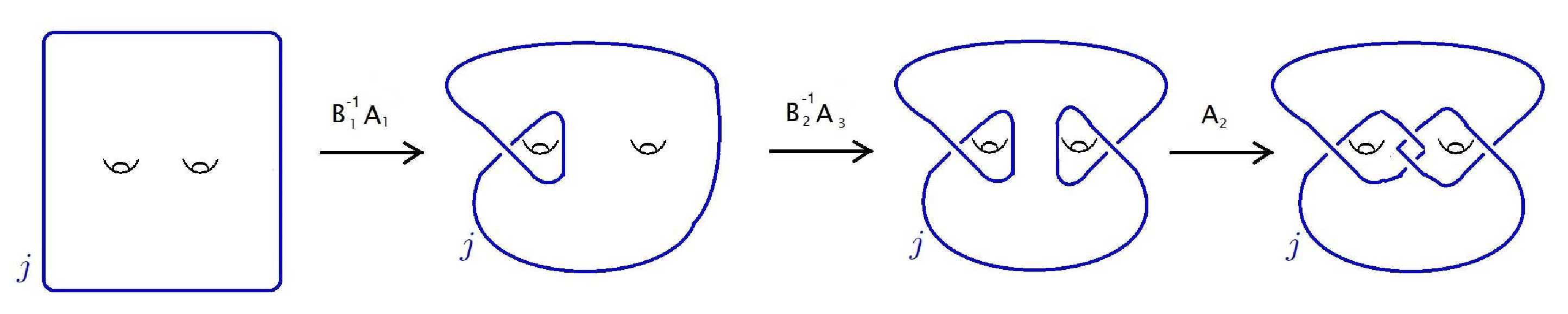}
\end{center}
\caption{Using the genus 2 automorphisms to wind an unknot into a figure eight.}
\label{figureeight}
\end{figure}

\begin{align}
{\cal O}_j\Big( \ \mbox{Pretzel}_{n_1 n_2 n_3} \ \Big) = U \ {\cal O}^{(2)}_j \ U^{-1}, \ \ \ U =  A_1^{m_1} A_2^{m_2} A_3^{m_3} \ B_2^{-1} A_3 B_1^{-1} A_1
\end{align}
\smallskip\\
where $(n_1, n_2, n_3) = (2 m_1 + 1, 2 m_2, 2 m_3 + 1)$. This 3-parametric family includes many quite non-trivial knots, and will be the main playground in the present paper. Using level $K$ refined TQFT representations, we straightforwardly find

\begin{align}
\nonumber \dfrac{ {\cal Z}_1( \ \mbox{Pretzel}_{1 2 1} = 4_1 \ ) }{ {\cal Z}_1( \ \mbox{Pretzel}_{1 0 1} = \bigcirc \ ) } \ = \ & q^{-1} + q^{K/2 - 1} - q^{K - 2} + q^{K - 1} - q^{3K/2 - 2} \\
\ = \ & t^{-3} q^{-2} \ \big( 1 - t + t q - t^2 q + t^3 q \big)
\end{align}
\smallskip\\
This is the same procedure that has been used in \cite{AS}, only now in genus 2 setting. It is straightforward and quite fast: more examples are provided in the next section.

\paragraph{$4_1$ and its mirror.} We find $Z(4_1; q,t) = 1 - t + t q - t^2 q + t^3 q$ as the refined amplitude for the figure eight knot $4_1$. It is easy to check that for the mirror $4_1$ on the genus two surface, which is obtained by inverting $A_i \rightarrow A_i^{-1}, B_i \rightarrow B_i^{-1}$ the answer would be $Z(\overline{4_1}; q,t) = Z(4_1; q^{-1},t^{-1}) = 1 - t^{-1} + t^{-1} q^{-1} - t^{-2} q^{-1} + t^{-3} q^{-1}$, which is not the same. Since $4_1$ and its mirror are identical, this implies that the amplitude $Z$ can not literally be a knot invariant -- but its norm $Z(K; q,t) Z(K; q^{-1},t^{-1})$ could.

\pagebreak

\section{Refined Chern-Simons Invariants of Pretzel knots}

In this section we provide more examples of refined knot amplitudes in genus 2. With the full machinery of the mapping class group at hand, one can compute refined Chern-Simons amplitudes for any knot in genus 2. For illustration, we present a detailed exposition for the Pretzel knots.

\paragraph{}Note that 3-index Pretzel knots possess both cyclic $(n_1,n_2,n_3) \simeq (n_2,n_3,n_1)$ and reversal $(n_1,n_2,n_3) \simeq (n_3,n_2,n_1)$ symmetries, hence, they are completely symmetric. All knot amplitudes are normalized by the amplitude of the unknot, and further normalized to be a polynomial in non-negative powers of $q^{\geq 0} t^{\geq 0}$, starting with 1.

\paragraph{}Let us start by gradually increasing $n$'s in the small positive area, keeping the color $j = 1$. This gives a bunch of simple knots from the Rolfsen table:

\begin{align*}
\begin{array}{c|c|ccc}
(n_1,n_2,n_3) & \mbox{ Knot } K & \mbox{ Normalized Amplitude } {\cal Z}_1(K) / {\cal Z}_1( \bigcirc )
\\ \hline & &
\\
(1,2,1) & 4_1 & \begin{array}{ll} 1-t+t q-t^2 q+t^3 q \\ \emph{} \end{array}
\\ \hline & &
\\
(1,4,1) & 6_1 & \begin{array}{ll} 1-t+t q-2 t^2 q+t^2 q^2+t^3 q-t^3 q^2+t^4 q^2 \\ \emph{} \end{array}
\\ \hline & &
\\
(1,2,3), (3,2,1) & 6_2 & \begin{array}{ll} 1 - t + 2 t q - 2 t^2 q + t^3 q + t^2 q^2 - 2 t^3 q^2 + t^4 q^2 \\ \emph{} \end{array}
\\ \hline & &
\\
(1,6,1) & 8_1 &
\begin{array}{ll} 1-t+t q-2 t^2 q+t^2 q^2+t^3 q \\
- 2 t^3 q^2+t^3 q^3+t^4 q^2-t^4 q^3+t^5 q^3 \\ \emph{} \end{array}
\\ \hline & &
\\
(3,4,1), (1,4,3) & 8_4 & \begin{array}{ll} 1-t+2 t q-3 t^2 q+2 t^2 q^2+t^3 q \\ - 3 t^3 q^2+t^3 q^3+2 t^4 q^2-2 t^4 q^3+t^5 q^3 \\ \emph{} \end{array}
\\ \hline & &
\\
(1,2,5), (5,2,1) & 8_2 & \begin{array}{ll} 1-t+2 t q-2 t^2 q+2 t^2 q^2+t^3 q \\ -3 t^3 q^2+t^3 q^3+t^4 q^2-2 t^4 q^3+t^5 q^3 \\ \emph{} \end{array}
\\ \hline & &
\\
(3,2,3) & 8_5 & \begin{array}{ll} 1-t+3 t q-3 t^2 q+2 t^2 q^2+t^3 q \\ -4 t^3 q^2+t^3 q^3+2 t^4 q^2-2 t^4 q^3+t^5 q^3 \\ \emph{} \end{array}
\end{array}
\end{align*}

Note that some knots have two different genus 2 realizations, differing by a permutation of $n_1$ and $n_3$. The fact that the answers match provides a simple check of topological invariance of the refined TQFT. Continuing to 10 crossings,

\vspace{3ex}
\begin{align*}
\begin{array}{c|c|ccc}
(n_1,n_2,n_3) & \mbox{ Knot } K & \mbox{ Normalized Amplitude } {\cal Z}_1(K) / {\cal Z}_1( \bigcirc )
\\ \hline & &
\\
(1,8,1) & 10_1 & \begin{array}{ll} 1-t+t q-2 t^2 q+t^2 q^2 +t^3 q \\ -2 t^3 q^2+t^3 q^3 + t^4 q^2-2 t^4 q^3+ \\ t^4 q^4+t^5 q^3-t^5 q^4+t^6 q^4 \\ \emph{} \end{array}
\\ \hline & &
\\
(3,6,1), (1,6,3) & 10_4 & \begin{array}{ll} 1-t+2 t q-3 t^2 q+2 t^2 q^2+t^3 q \\ -4 t^3 q^2 +2 t^3 q^3+2 t^4 q^2-3 t^4 q^3 + \\ t^4 q^4 + 2 t^5 q^3-2 t^5 q^4+t^6 q^4 \\ \emph{} \end{array}
\\ \hline & &
\\
(1,4,5), (5,4,1) & 10_8 & \begin{array}{ll} 1 - t + 2 t q - 3 t^2 q + 3 t^2 q^2 + \\ t^3 q - 4 t^3 q^2 + 2 t^3 q^3 + 2 t^4 q^2 - 4 t^4 q^3 + \\ t^4 q^4+2 t^5 q^3-2 t^5 q^4+t^6 q^4 \\ \emph{} \end{array}
\\ \hline & &
\\
(3,4,3) & 10_{61} & \begin{array}{ll} 1-t+3 t q-4 t^2 q+3 t^2 q^2+t^3 q-\\5 t^3 q^2+2 t^3 q^3+3 t^4 q^2-4 t^4 q^3+\\t^4 q^4+2 t^5 q^3-2 t^5 q^4+t^6 q^4 \\ \emph{} \end{array}
\\ \hline & &
\\
(1,2,7), (7,2,1) & 10_2 & \begin{array}{ll} 1-t+2 t q-2 t^2 q+2 t^2 q^2+t^3 q \\ -3 t^3 q^2 + 2 t^3 q^3+t^4 q^2-3 t^4 q^3+ \\ t^4 q^4+t^5 q^3-2 t^5 q^4+t^6 q^4 \\ \emph{} \end{array}
\\ \hline & &
\\
(3,2,5), (5,2,3) & 10_{46} & \begin{array}{ll} 1-t+3 t q-3 t^2 q+3 t^2 q^2+t^3 q  \\ -5 t^3 q^2+2 t^3 q^3+2 t^4 q^2-4 t^4 q^3+ \\t^4 q^4+2 t^5 q^3-2 t^5 q^4+t^6 q^4 \\ \emph{} \end{array}
\end{array}
\end{align*}

\pagebreak

\begin{figure}[b]
\begin{center}
\includegraphics[width=0.5\textwidth]{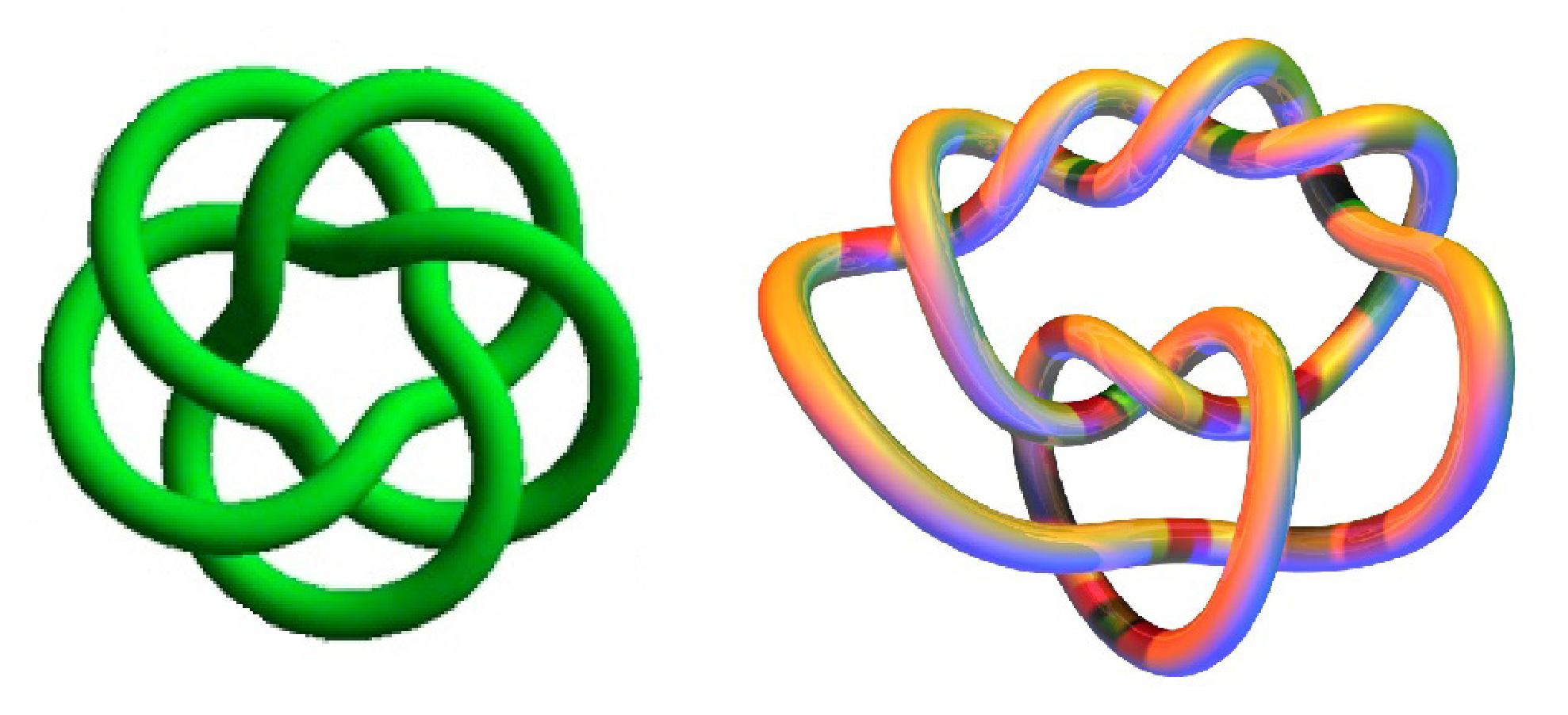}
\end{center}
\caption{Two realizations of the same knot $10_{124}$ -- in genus 1, as a torus knot $T_{3,5}$ and in genus 2, as a pretzel knot $(-2,3,5)$. The values of the amplitudes match.}
\end{figure}

\paragraph{}Another interesting series of examples, allowing to further test topological invariance, is obtained by allowing some of the indices $n_1,n_2,n_3$ to be negative or zero:

\begin{align*}
\begin{array}{c|c|ccc}
(n_1,n_2,n_3) & \mbox{ Knot } K & \mbox{ Normalized Amplitude } {\cal Z}_1(K) / {\cal Z}_1( \bigcirc )
\\ \hline & &
\\
(-1,0,3) & 3_1 & 1 + tq - t^2 q
\\ \hline & &
\\
(-3,0,-3) & 3_1 \ \# \ 3_1 & (1 + tq - t^2 q)^2
\\ \hline & &
\\
(3,0,-3) & 3_1 \ \# \ \overline{3_1} & (1 + tq - t^2 q)(1 - t - t^2 q)
\\ \hline & &
\\
(1,-2, 1) & 3_1 \simeq T_{2,3} & 1 + tq - t^2 q
\\ \hline & &
\\
(1,-2, 3) & 5_1 \simeq T_{2,5} & 1 + tq - t^2 q + t^2 q^2 - t^3 q^2
\\ \hline & &
\\
(1,-2, 5) & 7_1 \simeq T_{2,7} & 1 + tq - t^2 q + t^2 q^2 - t^3 q^2 + t^3 q^3 - t^4 q^3
\\ \hline & &
\\
(3,-2, 3) & 8_{19} \simeq T_{3,4} & 1+t q+t q^2-t^2 q-t^3 q^2
\\ \hline & &
\\
(3,-2, 5) & 10_{124} \simeq T_{3,5} & 1 + t q + t q^2 - t^2 q + t^2 q^3 - t^3 q^2 - t^3 q^3
\end{array}
\end{align*}
Two of the knots are composite -- $(-3,0,-3)$ a.k.a the \emph{Granny knot}, and $(3,0,-3)$ a.k.a. the \emph{square knot} -- they are connected sums (denoted $\#$) of trefoils. The refined amplitudes of these knots factorize, suggesting this is the general behaviour w.r.t. the connected sum operation. The others give alternative genus 2 realizations of torus knots, incluing the most complicated $(3,-2,3) = 8_{19} \simeq T_{3,4}$ and $(3,-2,5) = 10_{124} \simeq T_{3,5}$. The answers, that we obtain here with a genuinely genus 2 computation, match the corresponding results of the genus 1 computations of \cite{AS}. This provides a non-trivial check of topological invariance of the refined TQFT.

\pagebreak

As explained in \cite{noroots}, in refined Chern-Simons theory one can expect to unify all of the above examples into a single \emph{evolution} formula a-la \cite{evolution}, making the dependence on the winding numbers $m_1,m_2,m_3$ fully explicit\footnote{As a reminder, $(n_1, n_2, n_3) = (2 m_1 + 1, 2 m_2, 2 m_3 + 1)$.}. Let us briefly review here the argument of \cite{noroots}. First, by definition, the knot amplitude is given by

\begin{align}
{\cal Z}_j\big( \ \mbox{Pretzel}_{n_1,n_2,n_3} \ \big) = \langle 0, 0, 0 | \ I \ A_1^{-m_1} \ A_2^{-m_2} \ A_3^{-m_3} \ {\cal O}_j(\bigcirc) \ | 0, 0, 0 \rangle
\end{align}
\smallskip\\
Second, this formula can be expanded as a sum over intermediate states,

\begin{align}
{\cal Z}_j\big( \ \mbox{Pretzel}_{n_1,n_2,n_3} \ \big) = \sum\limits_{k_1,k_2,k_3} \ \Gamma^{(k_1,k_2,k_3)}_j \ T_{k_1}^{m_1} \ T_{k_2}^{m_2} \ T_{k_3}^{m_3}
\end{align}
\smallskip\\
where we used the fact that the $A$-twists are diagonal, and denoted

\begin{align}
\Gamma^{(k_1,k_2,k_3)}_j = \langle 0, 0, 0 | \ I \ | k_1,k_2,k_3 \rangle \cdot \langle k_1,k_2,k_3 \ | \ {\cal O}_j(\bigcirc) \ | 0, 0, 0 \rangle
\end{align}
\smallskip\\
Finally -- and this was the main point of \cite{noroots} -- knot operators in refined Chern-Simons theory are highly sparse. Even though \emph{a priori} the sum in the above formula goes over all $k_1,k_2,k_3$ in the admissible set, the matrix elements of knot operators, namely, $\Gamma^{(k_1,k_2,k_3)}_j$, are nonzero only for a few values of $k$, which are actually independent on $K$ at all. For example, in the fundamental case $j = 1$ these are

\begin{align}
\Gamma^{(2,2,0)}_1 = \Gamma^{(0,2,2)}_1 = \dfrac{t^2q-1}{tq(1-t)} \ \Gamma^{(0,0,0)}_1
\end{align}
\begin{align}
\Gamma^{(2,0,2)}_1 = \dfrac{(1-t^2q)(1-q)}{t q^2 (1-t)^2} \ \Gamma^{(0,0,0)}_1, \ \ \ \Gamma^{(2,2,2)}_1 = \dfrac{(1-t^2q)(1+tq)}{t^2q^2(1-t)} \ \Gamma^{(0,0,0)}_1
\end{align}
\smallskip\\
and all the other $\Gamma$'s vanish. This implies that

{\fontsize{10pt}{0pt}
\begin{align}
\nonumber & \dfrac{ {\cal Z}_1\big( \ \mbox{Pretzel}_{n_1,n_2,n_3} \ \big) }{ {\cal Z}_1\big( \ \mbox{Pretzel}_{1,0,1} \ \big) } = \dfrac{q^2t^2(1-t)^2}{(1-qt)^2} \ \left( 1 + \dfrac{t^2q-1}{tq(1-t)} \ \Big[ \ (qt)^{-m_1-m_2} + (qt)^{-m_2-m_3} \ \Big] + \right. \\ &
\left. + \dfrac{(1-t^2q)(1-q)}{t q^2 (1-t)^2} \ (qt)^{-m_1-m_3} + \dfrac{(1-t^2q)(1+tq)}{t^2q^2(1-t)} \ (qt)^{-m_1-m_2-m_3} \right)
\end{align}}
\smallskip\\
One can check that this, indeed, reproduces all the examples above. This seems to be a deformation of the formula of \cite{pretzels}, and it would be interesting to understand how to generalize this to Pretzel knots in higher genus, along the lines of \cite{pretzels}.

\section{The algebra of knot operators}

If one inserts one and the same knot several times, it is natural to expect that the result can be expressed as a linear combination of single insertions, summed over various colors. This implies that knot operators naturally form an algebra. In the usual Chern-Simons TQFT it was very simple, and looked like

\begin{align}
q = t: \ \ \ {\cal O}_i \big( \bigcirc \big) {\cal O}_j \big( \bigcirc \big) = \mathop{\sum\limits_{0 \leq k \leq K}}_{(i,j,k) \mbox{ admiss. }} {\cal O}_k \big( \bigcirc \big)
\end{align}
\smallskip\\
The refined knot operators, that we constructed above, enjoy a similar algebra:

\begin{align}
{\cal O}_i \big( \bigcirc \big) {\cal O}_j \big( \bigcirc \big) = g_i \ g_j \ \sum\limits_{0 \leq k \leq K} \ {\cal N}_{i j k} \ {\cal O}_k \big( \bigcirc \big)
\end{align}
\smallskip\\
One can think of this as a recursion relation, expressing knot operators with higher colors through the knot operators through lower colors. Solving it order by order, one finds completely explicit formulas

\begin{align}
& {\cal O}_0 = \big( {\cal O}_1 \big)^0 \ \equiv \ 1 \\
& \nonumber \\
& {\cal O}_1 = \big( {\cal O}_1 \big)^1 \\
& \nonumber \\
& {\cal O}_2 = \big( {\cal O}_1 \big)^2 - \dfrac{(1-q)(1+t)}{1-q t} \ \big( {\cal O}_1 \big)^0 \\
& \nonumber \\
& {\cal O}_3 = \big( {\cal O}_1 \big)^3 - \dfrac{(1-q)(2qt+q+t+2)}{1-q^2 t} \ \big( {\cal O}_1 \big)^1 \\
& \nonumber \ldots
\end{align}
\smallskip\\
expressing everything in terms of ${\cal O}_1$. It is not only easy to solve order by order, a general solution is not hard either, because the exact same algebra is satisfied by the Macdonald polynomials (this is one of the alternative definitions of ${\cal N}$, see \cite{AS}):

\begin{align*}
M_i\big( x, x^{-1} \big) M_j \big( x, x^{-1} \big) = g_i \ g_j \ \sum\limits_{0 \leq k \leq K} \ {\cal N}_{i j k} \ M_k \big( x , x^{-1} \big)
\end{align*}
\smallskip\\
This implies that knot operators ${\cal O}_j$ are recovered from the simplest knot operator ${\cal O}_1$ in the same way \footnote{In the torus case this fact was pointed out in \cite{GorskyNegut}. One can see that it is a very general fact.} as Macdonald polynomials $M_j(x,x^{-1})$ are recovered from the simplest Macdonald polynomial $M_1(x,x^{-1}) = x + x^{-1}$. The easiest way to do this recovery is to first express Macdonald polynomials through the Schur polynomials,

\begin{align*}
M_j\big( x, x^{-1} \big) = \sum\limits_{l = 0}^{[j/2]} \ \dfrac{q^{l} [j - 2 l + 1]}{[j - l + 1]} \ \prod\limits_{m = 0}^{l - 1} \dfrac{[j - l + 1 + m] [m + \beta - 1]}{[m + 1][j + \beta - 1 - m]} \ \chi_{j-2l}(x, x^{-1})
\end{align*}
\smallskip\\
and then express the Schur polynomials through the desired basis -- powers of $x+x^{-1}$:

\begin{align*}
\chi_{j-2l}\big( x, x^{-1} \big) = \sum\limits_{p = 0}^{[j/2]-l} \ \dfrac{(-1)^p (j - 2l - p)!}{p!(j - 2l - 2p)!} \ (x+x^{-1})^{j-2l-2p}
\end{align*}
\smallskip\\
Note, that the last formula is written in terms of the usual, not $q$-deformed, factorials. Putting these two together and replacing $x + x^{-1} \mapsto {\cal O}_1$, we find an explicit formula for all knot operators, colored by arbitrary representations $j$:

{\fontsize{8pt}{0pt}
\begin{align}
\boxed{ \ \ \ {\cal O}_j = \sum\limits_{l = 0}^{[j/2]} \ \dfrac{q^{l} [j - 2 l + 1]}{[j - l + 1]} \ \prod\limits_{m = 0}^{l - 1} \dfrac{[j - l + 1 + m] [m + \beta - 1]}{[m + 1][j + \beta - 1 - m]} \ \sum\limits_{p = 0}^{[j/2]-l} \dfrac{(-1)^p(j - 2l - p)!}{p!(j - 2l - 2p)!} \ \big( {\cal O}_1 \big)^{j-2l-2p} \ \ \ }
\label{General1}
\end{align}}
\smallskip\\
Note that this formula is completely general and applies to refined Chern-Simons TQFT in any genus, if it exists. The only external input, required by this formula, is the knowledge of the fundamental knot operator ${\cal O}_1$. Fortunately, we possess the duality definition eq. (\ref{duality}), which allows us to directly compute the genus-2 $\ {\cal O}_1$:

{\fontsize{9pt}{0pt}\begin{align}
\langle n+1, u+1, v | \ {\cal O}^{(1)}_1 \ | n, u, v \rangle = \dfrac{\left[ \frac{n+u+v}{2} + 1 + \beta \right] \ \left[ \frac{n+u-v}{2} + 1 \right] \ \big[ u + 2 \beta \big] \ \big[ n + 2 \beta \big] }{ \big[ u + \beta \big] \ \big[ u + \beta + 1 \big] \ \big[ n + \beta \big] \ \big[ n + \beta + 1 \big] }
\label{General2}
\end{align}
\begin{align}
\langle n+1, u-1, v | \ {\cal O}^{(1)}_1 \ | n, u, v \rangle = - \dfrac{\left[ \frac{u+v-n}{2} + 1 + \beta \right] \ \left[ \frac{u-v-n}{2} + 1 \right] \ \big[ n + 2 \beta \big] }{ \big[ u + 2 \beta - 1 \big] \ \big[ n + \beta \big] \ \big[ n + \beta + 1 \big] }
\label{General3}
\end{align}}
\smallskip\\
and all the other matrix elements vanish. Together, eqs. (\ref{General1}),(\ref{General2}),(\ref{General3}) give an explicit formula for all the $q,t$-deformed squares of $q$-6j symbols. This allows to define and study the algebra of knot operators for generic $q,t \in {\mathbb C}^{\star}$ without referring to TQFT or roots of unity, see \cite{genus2proof}.

\section{Discussion}

\vspace{2ex}

\paragraph{$\bullet$ Distinguishing mutants.} One of the most straightforward and interesting applications of refined Chern-Simons theory could be distinguishing mutants \cite{mutants} -- knots that cannot be distinguished by the usual Jones polynomials, or generally by HOMFLY polynomials colored by highest weights of symmetric or antisymmetric representations. Unfortunately, the knots that we have computed so far (the 3-index pretzels) do not have any non-trivial mutants among them. However, there might be such among the non-pretzel knots in genus 2, and it would be very interesting to check if refined Jones polynomials distinguish them or not. Another obvious possibility is to go to genus 3, where there exist non-trivial pretzel mutants.

\paragraph{$\bullet$ Higher genus.} The construction of present paper relies upon the Macdonald duality equation, that constrains the matrix elements of knot operators in genus 2. This duality equation is a deformation of the known Fourier duality equation for the squares of q-6j symbols. Following the same steps as we do in higher genus, one inevitably discovers that the matrix elements of knot operators are no longer degree 2 contractions of the q-6j symbols, but rather degree 4 contractions. This degree does not grow: for generic $g$ it stays degree 4. Genus 2 is a distinguished case from this point of view. To obtain a refined q,t-deformation of these degree 4 contractions, it is natural to look for degree 4 generalizations of Fourier/Macdonald duality; this remains to be done.

\paragraph{$\bullet$ Higher rank.} The main problem with generalization to higher rank is the fact that basis vectors in the vector space, associated to a surface of genus 2 (or higher), is no longer a decorated knot: it is a decorated trivalent graph. For a TQFT of type $A_n$, the decoration will include, as a part, assigning multiplicities of tensor products of representations to the trivalent intersections. It is not completely clear how this will affect the central identity of the present construction -- the Macdonald duality. In addition, introducing and handling multiplicities is simply very hard technically.

The usual solution to this problem is to only consider knots colored by the highest weights of symmetric or antisymmetric representations. This, however, does not seem to be possible within the mapping class group approach, since we do not choose which decorated graphs to include into the definition of the basis -- this is forced on us by the values of $N$ and $K$. The right methods and language to generalize to higher rank remain to be found.

\paragraph{$\bullet$ Higher genus DAHA's.} As discussed before in \cite{GorskyNegut}, knot operators in refined Chern-Simons theory on a torus generate an algebra which is isomorphic to the spherical DAHA, also known as the elliptic Hall algebra. Our results seem to suggest that a similar algebra exists in genus 2, generated by all knot insertion operators along all possible knots. In principle, using the formulas of present paper it should be possible to learn quite a lot about this algebra.

\paragraph{$\bullet$ Refined Chern-Simons as a two-parameter quantization.} It is known that knot operators in ordinary Chern-Simons theory provide a quantization of the Poisson algebra of functions on the moduli space of flat connections on the surface. The parameter $q$ plays the role of a quantum parameter, with $q \rightarrow 1$ being the classical limit, where the Poisson algebra is recovered. The fact that there exists a Macdonald q,t-deformation, with two independent "quantum" parameters, suggests that there exist two independent Poisson brackets for functions on the moduli space of flat connections. It would be interesting to make this and other statements about the "classical" limit of refined Chern-Simons theory more precise.

\paragraph{$\bullet$ Elliptic quantum groups.} One natural place where q,t-6j symbols with two deformation parameters appear in mathematical physics are the elliptic quantum groups, such as $U_{q,t}(sl_2)$ \cite{Felder}. However, these q,t-6j symbols also typically contain a third "spectral" or "dynamical" parameter, and satisfy a dynamical Yang-Baxter equation. The relation between elliptic quantum groups and refined Chern-Simons theory, if any, should involve a way to eliminate of the spectral parameter.

\paragraph{$\bullet$ Topological string theory.} Given a 3-manifold $M$, is known \cite{CSstring} that the partition function of topological string theory on $T^{\star} M$ agrees with the partition function of Chern-Simons theory on $M$. As explained in \cite{AS}, there is a refined version of this relation. Namely, for Seifert 3-manifolds $M$ (in particular, for $S^3$) Chern-Simons partition function can be refined, with the refinement following entirely from the action of the genus 1 mapping class group. The resulting partition function agrees with the partition function of the refined topological string on $T^{\star} M$ \cite{AS}. One can think of this relation as an alternative way to compute the refined topological string partition function on backgrounds of the form $T^{\star} M$. The results of present paper imply an extension of the class of manifolds that can be accessed this way, from Seifert to more general ones, constructed with the genus 2 mapping class group.

\section*{Acknowledgements}

We are indebted to G.Masbaum for enlightening explanations of the higher genus TQFT representations of mapping class groups. We are grateful to M.Aganagic, I.Cherednik, E.Gorsky, R.Kashaev, A.Morozov, N.Reshetikhin and C.Vafa for fruitful discussions. The work of S.A. was partly supported by the grants RFBR 15-01-04217 and 15-51-50034-YaF. The work of Sh.Sh. was partly supported by the grants RFBR 15-01-05990 and NSh-1500.2014.2.

\end{document}